\UseRawInputEncoding
\documentclass[a4paper,aps,prd,tightenlines,preprintnumbers,nofootinbib,showkeys, superscriptaddress,12pt]{revtex4-1}
\usepackage{fullpage}
\usepackage{amsfonts}
\usepackage{amsmath}
\usepackage{slashed}
\usepackage{amssymb}
\usepackage{graphicx}
\usepackage{makeidx}
\usepackage{cancel}
\usepackage{epic}
\usepackage{eepic}
\usepackage{epsfig}
\usepackage{latexsym}
\usepackage[dvipsnames]{xcolor}
\usepackage{float}
\usepackage{multirow}
\usepackage{ragged2e}
\usepackage[export]{adjustbox}
\usepackage{xurl,hyperref}
\usepackage{enumitem}
\hypersetup{colorlinks=true,citecolor=red,linkcolor=NavyBlue,urlcolor=NavyBlue}
\usepackage[utf8]{inputenc}
\usepackage[caption=false]{subfig}
 
\usepackage{natbib}
\usepackage{relsize}
\usepackage[left=2.2cm,right=2.2cm,top=2.cm,bottom=2.cm]{geometry}
\usepackage{mathptmx}
\begin{document}
\relscale{1.05}
\captionsetup[subfigure]{labelformat=empty}

\title{Revisiting the Scalar Leptoquark ($S_1$) Model with the Updated Leptonic Constraints}

\author{Bibhabasu De}
\email{bibhabasude@gmail.com}
\affiliation{Department of Physics, The ICFAI University Tripura, Kamalghat-799210, Tripura, India}

\date{\today}
%\preprint{IP-BBSR/2020-2}

\begin{abstract}
\noindent
The Standard Model, if extended to the energy scale of $\mathcal{O}(1)$ TeV, the known particle spectrum could be augmented with a scalar leptoquark. Within this minimally extended framework, explaining the anomalous magnetic moment and electric dipole moment simultaneously for the three lepton generations over a parameter space consistent with all the lepton flavor violating bounds is possible. Such a model can be tested or falsified through the collider search experiments and/or by probing the low-energy lepton phenomena. This work studies the current prospects of the model in the presence of recent experimental updates for the leptonic observables.
\end{abstract}
	
\maketitle	

\section{Introduction}
\label{sec:int}
\noindent
The Standard Model~(SM) has already explained the color and electroweak sectors up to a high degree of testable precision. Further, the discovery of the 125 GeV Higgs boson at the Large Hadron Collider~(LHC) has completed the proposed particle spectrum of the SM~\cite{ATLAS:2012yve,CMS:2012qbp}. However, certain experimental observations and theoretical issues can't be explained within the framework of the SM and thus indicate the presence of some New Physics~(NP) yet to be explored. For example, the idea of gauge coupling unification hints at a more fundamental theory corresponding to a single gauge group. The SM gauge group, i.e., $SU(3)_C\times SU(2)_L\times U(1)_Y$ can be considered as its effective low-energy version obtained via a particular symmetry-breaking chain. The list of such Grand Unified Theories~(GUT) includes $SU(4)$~\cite{Pati:1974yy}, $SU(5)$~\cite{Georgi:1974sy}, $SO(10)$~\cite{Georgi:1974my, Fritzsch:1974nn}, $E_6$~\cite{Kang:2007ib, Hati:2015awg}, etc. It is interesting to note that within a GUT structure, quarks and leptons can directly couple at the tree-level through a hypothetical mediator --- Leptoquark~(LQ)~(for recent reviews, see Refs.~\cite{Dorsner:2016wpm, Davidson:1993qk, Hewett:1997ce, Nath:2006ut}). Though, in principle, within a local quantum field theory LQs can either be scalar or vector, the scalar LQs are more useful to study the loop-induced Beyond Standard Model~(BSM) contributions~\cite{Blumlein:1996qp,Fajfer:2015ycq,Barbieri:2015yvd}. LQs are crucial from various phenomenological aspects. For example, an extension of the SM with a LQ can explain several B-meson anomalies~\cite{Dorsner:2013tla,Gripaios:2014tna,Becirevic:2015asa,Becirevic:2016yqi, Crivellin:2017zlb,Cline:2017aed,DiLuzio:2017chi,Mandal:2018kau,Aydemir:2019ynb,Crivellin:2019dwb,Asadi:2023ucx} or can contribute to the flavor violating processes like $\tau\rightarrow\mu\gamma$ and $h\rightarrow\tau\mu$~\cite{PhysRevD.93.015010}. LQs may also be significant for the dark matter phenomenology~\cite{Mandal:2018czf,Choi:2018stw,Mohamadnejad:2019wqb} and the production of scalar particles at the LHC~\cite{Bhaskar:2020kdr,Bhaskar:2022ygp,DaRold:2021pgn,Agrawal:1999bk,Enkhbat:2013oba}. Note that the simplest GUT extensions assume a heavy LQ~\cite{Super-Kamiokande:2014otb, Dorsner:2012nq} to evade the proton lifetime constraints, but they can't be produced at the LHC. However, there are GUT formulations that can explain the stability of proton with a TeV-scale scalar LQ~\cite{BUCHMULLER1986377,Murayama:1991ah,Dorsner:2005fq, GEORGI1979297, FileviezPerez:2007bcw, Senjanovic:1982ex}. Thus, in this paper, the later GUT motivation will be considered as the gauge theoretical background for the new interactions, i.e., the SM will be extended to an energy scale of $\mathcal{O}(1)$ TeV to augment the observed particle spectrum with a scalar LQ.  

Recent experiments have resulted in some remarkable observations in the lepton sector, which may indicate towards a possible BSM theory yet to be discovered. For example, in 2021 a combined result from the Fermilab-based Muon $g-2$ collaboration and Brookhaven National Laboratory~(BNL) showed a $4.2\sigma$ discrepancy between the predicted and measured values of the anomalous magnetic moment of muon~\cite{Abi:2021gix,Albahri:2021ixb}. The result has been updated very recently on August 2023, enhancing the significance to $5\sigma$~\cite{Muong-2:2023cdq}\,\footnote{A recent lattice calculation of the hadronic vacuum polarization~(HVP) term by the BMW collaboration~\cite{Borsanyi:2020mff} and a preliminary experimental update from the CMD-3 detector~\cite{CMD-3:2023alj} indicate a significant tension with the present data which may result in a smaller and less significant  discrepancy~\cite{Colangelo:2022jxc} between the predicted and observed values of $(g-2)_\mu$.}. Moreover, a precision measurement of the fine-structure constant using either Cesium~(Cs)~\cite{Parker_2018} or Rubidium~(Rb)~\cite{Morel:2020dww} indicates a similar anomaly in $(g-2)_e$. However, note that a relative sign between the two results leads to an experimental dispute that can't be settled with the present technologies. LQs can play a vital role in explaining the discrepancy in $(g-2)_\mu$~\cite{Djouadi1990,PhysRevD.53.555,Cheung:2001ip,Dorsner:2019itg,Greljo:2021xmg,Kowalska:2018ulj,Athron:2021iuf}. Moreover, in the presence of a scalar LQ, various NP signatures, e.g., the neutrino oscillation, $W$ mass anomaly, lepton flavor violating decays and dark matter can be connected to the $(g-2)_{e,\,\mu}$ anomalies within a single BSM formulation~\cite{Chen:2022hle,Choi:2018stw,Saad:2020ihm, Gherardi:2020qhc,Chen:2022hle,Bhaskar:2022vgk,Parashar:2022wrd,Zhang:2021dgl,ColuccioLeskow:2016dox,Mandal:2019gff,Ghosh:2022vpb}. LQs can also have important implications to explain the electric dipole moment~(EDM) of leptons~\cite{Altmannshofer:2020ywf, Dekens:2018bci, Fuyuto:2018scm}.

In this paper, a {\it minimal} extension of the SM has been considered with a scalar LQ $S_1(\mathbf{\bar{3}},\,\mathbf{1},\,1/3)$ at an energy scale of $\mathcal{O}(1)$ TeV. In Refs.~\cite{Mandal:2019gff,Bigaran:2021kmn}, it has already been studied in detail that such a simple BSM framework can easily explain all the possible NP signatures and experimental constraints in the lepton sector. However, we shall see that the scenario could be simplified further if formulated with a particular flavor ansatz. The present work will try to constrain the parameter space for all the three lepton generations simultaneously considering the current experimental updates on $(g-2)_\ell$ and EDM. However, due to experimental inadequacy, the $\tau$-sector is not at all interesting compared to $e$ and $\mu$. For $e$-sector, both experimental possibilities~(i.e., the results from the Cs and Rb experiments) will be addressed through a common generic formulation. A direct consequence of augmenting the SM with a LQ is opening up the 2-body and 3-body charged lepton flavor violating~(CLFV) decay channels and initiating a possibility for the lepton flavor violating Higgs decays~\cite{PhysRevD.93.015010, Chang:2016zll, Husek:2021isa, DelleRose:2020qak}. However, the experimental upper limits associated with the non-observation of these processes can easily be explained within the considered model by adjusting the lepton-quark couplings in a $3\times 3$ flavor basis, making the parameter space consistent with the CLFV bounds. The paper has been organized as follows. Sec.~\ref{sec:model} introduces the new interactions arising at the TeV scale. In Sec.~\ref{sec:bounds}, $(g-2)_\ell$ and EDM have been defined along with their recent experimental bounds. Sec.~\ref{sec:anal} elaborates on the one-loop BSM contributions to the $\ell\ell\gamma$ vertex appearing in the presence of $S_1$, whereas in Sec.~\ref{sec:num}, the allowed parameter space has been analyzed using numerical techniques. Finally, the outcomes have been summarized in Sec.~\ref{sec:conc}.

\section{The Model: A Minimal Extension of the SM}
\label{sec:model}
\noindent
The considered model assumes a simple extension of the SM at a NP scale $\Lambda\sim\mathcal{O}(1)$ TeV, where the known particle spectrum gets augmented with a scalar Leptoquark~(LQ) of electromagnetic~(EM) charge $1/3$ --- usually labeled as $S_1\equiv S_1(\mathbf{\bar{3}},\,\mathbf{1},\,1/3)$. Following the notations of Ref.~\cite{Dorsner:2016wpm}, the NP Lagrangian can be cast as,
\begin{align}
\mathcal{L}_{\Lambda}&=\Big[\lambda_L^{ij}\left(\bar{Q}_L^{Cia\beta}\epsilon^{ab}L_L^{jb}\right)S^\beta_1+\lambda_R^{ij}\left(\bar{u}_R^{Ci\beta}S^\beta_1\ell_R^{j}\right)+{\rm h.c.}\Big]+\bar{M}_{S_1}^2(S_1^\dagger S_1)+\kappa(H^\dagger H)(S_1^\dagger S_1),\nonumber\\
&=\Bigg[\Big\{\bar{u}_L^{Ci\beta}\left(\mathbb{V}^\dagger\lambda_L\right)^{ij}\ell_L^{j}-\bar{d}_L^{Ci\beta}\lambda_L^{ij}\nu_L^{j}\Big\}S^\beta_1+\lambda_R^{ij}\left(\bar{u}_R^{Ci\beta}S^\beta_1\ell_R^{j}\right)+{\rm h.c.}\Bigg]\nonumber\\
&\qquad\qquad\qquad\qquad\qquad\qquad\qquad\qquad\qquad\qquad+\bar{M}_{S_1}^2(S_1^\dagger S_1)+\kappa(H^\dagger H)(S_1^\dagger S_1).
\label{eq:Lag_1}
\end{align}
Here, the EM cahrge has been defined as $Q_{\rm EM}=T_3+Y$. $Q_L\equiv (u_L\quad d_L)^T$ and $L_L\equiv(\nu_L\quad \ell_L)^T$ denote the left-handed quark and lepton doublets, whereas $u_R$ and $\ell_R$ stand for the right-handed up-type quarks and charged leptons, respectively. The superscript $C$ defines the charge conjugation. The indices $\{i,\,j\}$ and $\{a,\,b\}$ define the flavor and $SU(2)$ indices, respectively. $\beta$ refers to the color index and $\mathbb{V}$ defines the CKM matrix. Eq.~\eqref{eq:Lag_1} assumes the down-type quark and charged lepton Yukawas to be in the physical basis. Since neutrinos are insignificant for the low-energy phenomenology, the PMNS matrix has been set to identity. After electroweak symmetry breaking~(EWSB) only the SM Higgs acquires a vacuum expectation value~(VEV) as, 
\begin{align}
H=\frac{1}{\sqrt{2}}\left(\begin{array}{c}
0\\
v+h
\end{array}\right),
\end{align}
where $v=246$ GeV. Thus, the physical mass of $S_1$ can be cast as,
\begin{align}
M_{S_1}=\sqrt{\bar{M}_{S_1}^2+\frac{\kappa v^2}{2}}~,
\end{align}
where, $\bar{M}_{S_1}$ is the bare mass term and $\kappa$ is a dimensionless coupling. In principle, one should consider the kinetic term for $S_1$ in Eq.~\eqref{eq:Lag_1}. However, the NP contributions arising through the interaction of $S_1$ with the gauge bosons~(gluon and photon to be particular)~\footnote{For a detailed study, see Ref.~\cite{Bhaskar:2020kdr}.} are irrelevant in the lepton sector. Hence, the kinetic term can be dropped for simplicity. 

The NP couplings $\lambda_{L,R}^{ij}$ play a crucial role in describing the low-energy lepton phenomena. At this point, one can easily rotate the Yukawa matrix to the physical basis and constrain the parameter space through the leptonic observables. However, the computational rigor can be reduced through a careful analysis of $\Delta a_\ell=(g-2)_\ell/2$ and charged lepton flavor violating~(CLFV) processes. In Sec.~\ref{sec:anal}, we shall see that $\Delta a_\ell$ can be decomposed into two terms --- chirality-conserving and chirality-flipping. The former contribution is suppressed by $m_\ell^2$ whereas the latter is proportional to the mass of the virtual fermion~(here, the SM quarks) appearing in the loop~[see Fig.~\ref{fig:loops}]. Therefore, the largest NP contribution to $\Delta a_\ell$ corresponds to the $t$-quark loop, and within the perturbative regime of Yukawa couplings, one can easily neglect the $u$ and $c$ quark contributions to $\Delta a_\ell$ considering the mass hierarchy among the three quark generations. Thus, to a good approximation, the mixing among the quarks can be ignored. 

Following the above discussion, one may be tempted to assume a minimal flavor structure for enhancing the loop contribution to $\Delta a_\ell$~($\ell=e,\, \mu,\, \tau$) as follows:
\begin{align}
\lambda_{L,R}=\left(\begin{array}{c c c}
0 & 0 & 0 \\
0 & 0 & 0 \\
\lambda_{L,R}^{t,e} & \lambda_{L,R}^{t,\mu} & \lambda_{L,R}^{t,\tau}
\end{array}\right).
\label{eq:flav1}
\end{align}

However, it can be readily understood that the parameter space presented in Eq.~\eqref{eq:flav1} will be strongly constrained through the 2-body and 3-body lepton flavor violating decays. For example, if one sets $|\lambda_{L,R}^{t,e}|\sim \mathcal{O}(1)$, BR$(\mu\to e\gamma)<4.2\times 10^{-13}$~\cite{MEG:2016leq} leads to an upper limit $|\lambda_{L,R}^{t,\mu}|<10^{-8}$, making it impossible to explain $\Delta a_\mu$ within the assumed parameter space. A similar argument goes for the $\tau$-sector. Therefore, the minimal flavor ansatz should be so chosen that it can maximize the NP contribution to $\Delta a_\ell$ while explaining the non-observation of all the CLFV processes in the most economical way. Eq.~\eqref{eq:flav} represents the {\it minimal} Yukawa structure for this simplified model.
%However, to explain the observed discrepancies in the anomalous magnetic moments and the discovery prospects of the electric dipole moments for all the three lepton generations~($e,\,\mu,\,\tau$), along with the non-observation of the charged lepton flavor violating~(CLFV) processes, e.g., $\ell_j\to \ell_i\gamma$~\cite{MEG:2016leq, Baldini:2013ke, Aushev:2010bq, BaBar:2009hkt}, $\ell_j\to 3\ell_i$~\cite{Blondel:2013ia, BELLGARDT19881,Aushev:2010bq,Hayasaka:2010np} and $h\to \ell_i\ell_j$~\cite{ATLAS:2019old, ATLAS:2019pmk, CMS:2017con}, it is helpful to consider the following {\it minimal} flavor ansatz:
\begin{align}
\lambda_{L,R}=\left(\begin{array}{c c c}
0 & 0 & \lambda_{L,R}^{u,\tau} \\
0 & \lambda_{L,R}^{c,\mu} & 0 \\
\lambda_{L,R}^{t,e} & 0 & 0
\end{array}\right).
\label{eq:flav}
\end{align}
For Eq.~\eqref{eq:flav} one could have equivalently chosen the diagonal line, i.e., $\lambda_{L,R}={\rm diag}(\lambda_{L,R}^{u,e},\, \lambda_{L,R}^{c,\mu},\, \lambda_{L,R}^{t,\tau})$. Though the phenomenology of $\mu$ and $\tau$-sector would remain mostly unchanged, but due to the $u$-quark mass suppression, this diagonal Yukawa structure would lead to non-perturbative values of $|\lambda_{L,R}^{u,e}|$ for explaining the observed discrepancy in $(g-2)_e$. Note that, the zeros in Eq.~\eqref{eq:flav} are completely from the phenomenological perspective. 
\section{New Physics Observables and Experimental Bounds}
\label{sec:bounds}
\begin{figure}[!ht]
\centering
\includegraphics[scale=0.7]{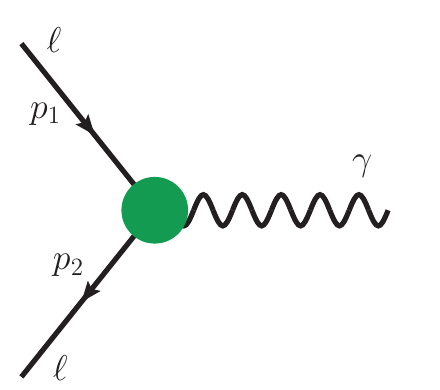}
\caption{Effective $\ell\ell\gamma$ vertex. $p_{1,\,2}$ represent the external momenta, with $q=p_1-p_2$ being the photon momentum.}
\label{fig:vert}
\end{figure}

\noindent
The most generic gauge invariant representation for the effective $\ell\ell\gamma$ vertex corresponding to Fig.~\ref{fig:vert} is given by, 
\begin{align}
\Gamma^\mu_{\ell\ell\gamma}= \gamma^\mu \mathcal{F}_1(q^2)+\left\{i\mathcal{F}_2(q^2)+\mathcal{F}_4(q^2)\,\gamma^5\right\}\left(\frac{\sigma^{\mu\nu}q_\nu}{2m_\ell}\right)+\mathcal{F}_3(q^2)(q^\mu\slashed{q}-q^2\gamma^\mu)\gamma^5,
\label{eq:loop_pht}
\end{align}
where $\mathcal{F}_{(1,2,3,4)}$ are the form factors and $q$ represents the photon momentum. 
However, in the case of an off-shell photon, there should be additional contributions in Eq.~\eqref{eq:loop_pht}. Note that the form factors $\mathcal{F}_3$ and $\mathcal{F}_4$ must vanish in any parity-conserving theory~(e.g., QED) and can only arise through the diagrams where electroweak~(EW) gauge bosons appear as the virtual particles. Thus, the renormalized vertex correction in QED results in~\cite{PhysRev.73.416},
\begin{align}
\mathcal{F}_1(0)=0,\qquad\mathcal{F}_2(0)=\frac{\alpha_{\rm EM}}{2\pi},
\end{align}
where $\mathcal{F}_1(0)$ corresponds to the correction in EM charge while $\mathcal{F}_2(0)$ represents the QED contribution to the anomalous magnetic moment of leptons at $\mathcal{O}(\alpha_{\rm EM})$. However, in the presence of the weak gauge bosons, the $\ell\ell\gamma$ vertex gets modified as~\cite{Hollik:1991qb}\footnote{The axial vector coupling associated with the form factor $\mathcal{F}_3$ vanishes for
on-shell photons as a consequence of the Ward identity~\cite{Ward:1950xp}.},
\begin{align}
\Gamma^\mu_{\rm EW}&=e\Bigg[\left(1+\frac{\delta e}{e}\right)\gamma^\mu + \frac{i\sigma^{\mu\nu}q_\nu}{2m_\ell}\left\{\frac{\alpha_{\rm EM}}{2\pi}+\mathcal{F}^{EW}_2(0)\right\}+\frac{\sigma^{\mu\nu}q_\nu}{2m_\ell}\gamma^5\mathcal{F}_4(0) \Bigg],
\label{eq:QED7}
\end{align} 
where $\delta e$ denotes the sum of the charge correction at one-loop order and the corresponding counterterm. The additional contribution to the anomalous magnetic moment can be parametrized as~\cite{Jackiw:1972jz},
\begin{align}
\mathcal{F}_2^{EW}(0)=\frac{\mathcal{G}_F\, m_\ell^2}{8\sqrt{2}\,\pi^2}\left[\frac{5}{3}+\frac{1}{3}(1-4\sin^2\theta_W)^2+\mathcal{O}\left(\frac{m_\ell^2}{M_W^2}\right)\right],
\end{align}
where, $\mathcal{G}_F$, $\theta_W$, and $M_W$ signify the Fermi constant, weak mixing angle, and mass of the $W$-boson, respectively. Moreover, considering the leading order~(LO) hadronic contribution, one obtains~\cite{Gourdin:1969dm},
\begin{align}
    \mathcal{F}_2(0)^{\rm Had}{\rm [LO]}=\left(\frac{\alpha_{\rm EM}}{\pi\sqrt{3}}\right)^2\int_{m^2_\pi}^\infty\frac{K(s)}{s}R^{(0)}(s)\,ds,
\end{align}
where, $K(s)$ stands for the QED kernel function~\cite{Brodsky:1967sr} and $R^{(0)}(s)$ represents the ratio of electron-positron bare annihilation cross into the hadrons to the cross section of muon-pair production with center of mass energy $\sqrt{s}$. However, this leading order hadronic contribution $\mathcal{F}_2(0)^{\rm Had}{\rm [LO]}$ includes a significant amount of uncertainty which might be resolved soon through the updated lattice calculations~\cite{Borsanyi:2020mff}. 

The last term in Eq.~\eqref{eq:QED7}, i.e., $\mathcal{F}_4(0)$ represents the leading order SM contribution to the electric dipole moment~($d_\ell$) of leptons. As already mentioned in Sec.~\ref{sec:int}, despite considering all the SM contributions these leptonic observables exhibit a sharp discrepancy with the experimental results. 
\subsection{Anomalous Magnetic Moment}
The best available SM prediction for the anomalous magnetic moment of muon is given by $a_\mu^{\rm SM}= 116 591 810(43)\times 10^{-11}$~\cite{Aoyama:2020ynm}, whereas the recent experimental data from Muon $g-2$ collaboration results in a world average of $a_\mu^{\rm Exp}=116 592 059(22)\times 10^{-11}$~\cite{Muong-2:2023cdq}, leading to a discrepancy, 
\begin{align}
\Delta a_\mu= a^{\rm Exp}_\mu- a^{\rm SM}_\mu=(2.49 \pm 0.48) \times 10^{-9}~(5.0\,\sigma).
\end{align}
As discussed, it can be one of the most remarkable signatures of a possible BSM sector. Further, in the context of electrons, experiments indicate a similar anomaly in $(g-2)_e$. A precision measurement of the fine structure constant through the recoil of ${\rm Cs}^{133}$ atoms has yielded a notable contradiction between the measured and predicted values of $a_e$ as~\cite{Parker_2018}, 
\begin{align}
\Delta a^{\rm (Cs)}_e =a^{\rm Exp\,(Cs)}_e-a^{\rm SM}_e= (-8.8 \pm 3.6) \times 10^{-13}~(2.4\,\sigma).
\label{eq:e_Cs}
\end{align}
However, for the same $ a_e$ a Rubidium based experiment results in~\cite{Morel:2020dww},
\begin{align}
\Delta a^{\rm (Rb)}_e =a^{\rm Exp\,(Rb)}_e-a^{\rm SM}_e= (4.8 \pm 3.0) \times 10^{-13}~(1.6\,\sigma).
\label{eq:e_Rb}
\end{align} 
Note that, despite having a significant expectation value, the experimental measurements for $\Delta a_e$ have large error bars. However, this paper is able to address both of the results for $\Delta a_e$, along with the non-zero value of $\Delta a_\mu$ within a common BSM framework. 

Unlike the first two generations, measuring the anomalous magnetic moment of $\tau$ is extremely challenging due to its short lifetime. Thus, $a_\tau^{\rm Exp}$ can only be traced back from the secondary particles produced through the decay of $\tau$. The latest experimental bound~(95\% CL) can be quoted as~\cite{DELPHI:2003nah, ParticleDataGroup:2022pth},
\begin{align}
-0.052<a_\tau<0.013,
\end{align} 
whereas, the corresponding SM prediction is given by, $a_\tau^{\rm SM}=117721(5)\times 10^{-8}$~\cite{Eidelman:2007sb}. 
\subsection{Electric Dipole Moment}
The precision measurement of the electric dipole moment of leptons can be crucial to search for the NP. EDM can be related to the form factor $\mathcal{F}_4$ as, $d_\ell=e\mathcal{F}_4(0)/m_\ell$, for which the SM predicts $|\mathcal{F}^e_4(0)|<|\mathcal{F}^\mu_4(0)|<|\mathcal{F}^\tau_4(0)|\approx 10^{-23}$~\cite{Booth:1993af,PhysRevD.54.3377, PhysRevLett.125.241802, PhysRevD.103.013001}, i.e., $|d_\tau^{\rm SM}|\simeq 10^{-37}~e\,{\rm cm}$. It is much smaller than the experimental sensitivity. Thus, any observation of lepton EDM can be treated as a direct evidence of some New Physics interaction. The experimental upper limits for the three lepton generations can be read as~\cite{ACME:2018yjb, Muong-2:2008ebm, Belle:2002nla},
\begin{align}
&|d_e|<0.11\times 10^{-28}~e\,{\rm cm}~(90\%~{\rm CL}),\nonumber\\
&|d_\mu|<1.8\times 10^{-19}~e\,{\rm cm}~(95\%~{\rm CL}),\nonumber\\
&{\rm Re}\,[d_\tau]\supset\left[-0.220,\, 0.45\right] \times 10^{-16}~e\,{\rm cm}~(95\%~{\rm CL}),\nonumber\\
&{\rm Im}\,[d_\tau]\supset\left[-0.250,\, 0.08\right] \times 10^{-16}~e\,{\rm cm}~(95\%~{\rm CL}).
\end{align} 
These experimental bounds on $\Delta a_\ell$ and $d_\ell$ will be simultaneously considered to constrain the chosen parameter space for each lepton generation. 
\subsection{CLFV Processes}
In general, the CLFV decays are allowed in a $S_1$-LQ extension of the SM. However, there is no positive signal from the ongoing experiments~\cite{MEG:2016leq, Baldini:2013ke, Aushev:2010bq, BaBar:2009hkt,Blondel:2013ia, BELLGARDT19881,Aushev:2010bq,Hayasaka:2010np,ATLAS:2019old, ATLAS:2019pmk, CMS:2017con} supporting the lepton flavor violating processes and thus only leads to upper bounds on the Yukawa couplings. Therefore, the non-observation of the 2-body and 3-body CLFV decays can easily be accommodated in this considered model if one follows the Yukawa structure defined by Eq.~\eqref{eq:flav} without any conflict with the experimental data. Thus, the minimal parameter space chosen here is automatically consistent with all the CLFV bounds. 
\section{BSM Contributions to $(g-2)_\ell$ and EDM}
\label{sec:anal}
\begin{figure}[!ht]
\centering
\subfloat[(a)]{\includegraphics[scale=0.6]{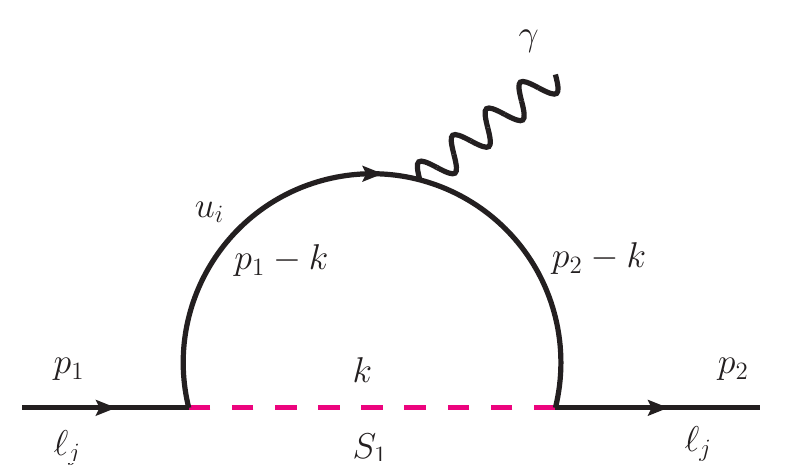}}\,\,
\subfloat[(b)]{\includegraphics[scale=0.6]{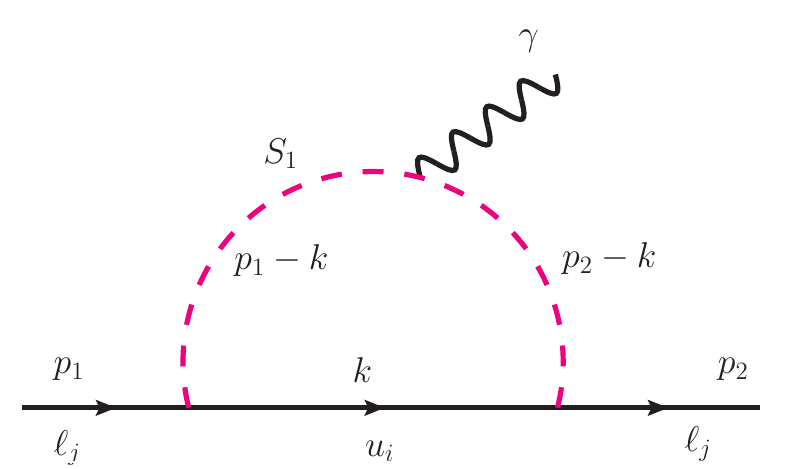}}
\caption{BSM contributions to the $\ell\ell\gamma$ vertex, where (a) the up-type quarks couple to the photon~(Type-1 diagram), and (b) the LQ $S_1$ couples to the photon~(Type-2 diagram). $p_1,\,p_2$ represent the external momenta.}
\label{fig:loops}
\end{figure}

\noindent
As already stated in Sec.~\ref{sec:int}, in the presence of a scalar LQ, there can be new contributions to the $\ell\ell\gamma$ vertex at one-loop order. Fig.~\ref{fig:loops}(a) shows the case where the photon couples to the up-type quarks, while Fig.~\ref{fig:loops}(b) represents the situation when photon touches the $S_1$ propagator~(magenta line). The former will be referred to as Type-1 diagram, while the latter will be called Type-2 for convenience.
\subsection{Type-1 Diagram} The correction term to $\ell_j\ell_j\gamma$ vertex due to the Type-1 diagram can be computed as,
\begin{align}
\Delta \Gamma^\sigma_1&= iN_C\int\frac{d^4k}{(2\pi)^4}\Bigg[(-\lambda_L^{ij}P_L+\lambda_R^{ij}P_R)\frac{(\slashed{p}_2-\slashed{k}+m_i)}{(k-p_2)^2-m_i^2}(Q_{\rm EM}^i\gamma^\sigma)\frac{(\slashed{p}_1-\slashed{k}+m_i)}{(k-p_1)^2-m_i^2}\nonumber\\
&\qquad\qquad\qquad\qquad\qquad\qquad\qquad\qquad\qquad\times\frac{1}{k^2-M_{S_1}^2}\{-(\lambda_L^{ij})^*P_R+(\lambda_R^{ij})^*P_L\}\Bigg]\nonumber\\
&\equiv iQ_{\rm EM}^i\,N_C\int\frac{d^4k}{(2\pi)^4}\Bigg[\frac{\mathcal{N}_1^\sigma}{\mathcal{D}_1}\Bigg].
\label{eq:t1}
\end{align}
Here $N_C=3$ defines the color degeneracy factor, and $Q_{\rm EM}^i=2/3$ represents the EM charge of up-type quarks in the unit of electronic charge $e$. $m_i$ denotes the up-type quark masses for $i=u,\,c,\,t$. The numerator can be rearranged as,
\begin{align}
\mathcal{N}_1^\sigma=\frac{1}{2}\Bigg[\mathcal{A}_1\Big\{&(\slashed{p}_2-\slashed{k})\gamma^\sigma(\slashed{p}_1-\slashed{k})+m_i^2\gamma^\sigma\Big\}+\mathcal{A}_2m_i\Big\{(\slashed{p}_2-\slashed{k})\gamma^\sigma+\gamma^\sigma(\slashed{p}_1-\slashed{k})\Big\}\nonumber\\
&+ \mathcal{A}_3\gamma^5\Big\{(\slashed{p}_2-\slashed{k})\gamma^\sigma(\slashed{p}_1-\slashed{k})+m_i^2\gamma^\sigma\Big\}+\mathcal{A}_4\,m_i\gamma^5\Big\{(\slashed{p}_2-\slashed{k})\gamma^\sigma+\gamma^\sigma(\slashed{p}_1-\slashed{k})\Big\}\Bigg],
\end{align}
where,
\begin{align}
\mathcal{A}_1&=|\lambda_R^{ij}|^2+|\lambda_L^{ij}|^2\,,\qquad\qquad\mathcal{A}_2=-2\,{\rm Re} [(\lambda_L^{ij})^*\lambda_R^{ij}],\nonumber\\
\mathcal{A}_3&=|\lambda_R^{ij}|^2-|\lambda_L^{ij}|^2\,,\qquad\qquad\mathcal{A}_4=-2\,{\rm Im} [(\lambda_L^{ij})^*\lambda_R^{ij}].
\end{align}
After Feynman parametrization, the denominator can be cast as,
\begin{align}
\mathcal{D}_1=n^2-\Delta_1(x),
\end{align}
where $n=k-yp_1-zp_2$ and $\Delta_1(x)=M_{S_1}^2\Big[x+\rho_i(1-x)\Big]$. $x,\,y,\,z$ are the Feynman parameters and $\rho_i=(m_i/M_{S_1})^2$. This calculation assumes an on-shell photon and the physically viable approximation of $(m_\ell/M_{S_1})^2\to 0$. $m_\ell$ denotes the mass of the SM leptons. Integrating over the loop momentum $n$, the BSM contributions to the anomalous magnetic moment~($\Delta a^\ell_1$) and electric dipole moment~($d_1^\ell$) of the SM leptons can be defined as,
\begin{align}
\label{eq:a1}
\Delta a^\ell_1&=\frac{1}{8\pi^2}\Bigg[\mathcal{A}_1\left(\frac{m_\ell}{M_{S_1}}\right)^2G_1(\rho_i)+\mathcal{A}_2\left(\frac{m_\ell\,m_i}{M^2_{S_1}}\right)G_2(\rho_i)\Bigg],\\
d_1^\ell&=\frac{e}{8\pi^2}\Bigg[\mathcal{A}_3\left(\frac{m_\ell}{M^2_{S_1}}\right)G_1(\rho_i)+\mathcal{A}_4\left(\frac{m_i}{M^2_{S_1}}\right)G_2(\rho_i)\Bigg],
\label{eq:d1}
\end{align}
where, the functions $G_1$ and $G_2$ are given by,
\begin{align}
G_1(w)&=\int_0^1\left[\frac{x(1-x)^2}{x+(1-x)w}\right]\,dx=\frac{2+3w-6w^2+w^3+6w\,{\rm ln}\,w}{6(1-w)^4},\nonumber\\
G_2(w)&=\int_0^1\left[\frac{(1-x)^2}{x+(1-x)w}\right]\,dx=\frac{-3+4w-w^2-2\,{\rm ln}\,w}{2(1-w)^3}\,.
\end{align}

\subsection{Type-2 Diagram} Fig.~\ref{fig:loops}~(b) contributes to the $\ell_j\ell_j\gamma$ vertex as follows.
\begin{align}
\Delta \Gamma^\sigma_2&= iN_C\int\frac{d^4k}{(2\pi)^4}\Bigg[(-\lambda_L^{ij}P_L+\lambda_R^{ij}P_R)\frac{(\slashed{k}+m_i)}{k^2-m_i^2}.\frac{1}{(k-p_1)^2-M_{S_1}^2}\,.\,Q_{\rm EM}^{S_1}(p_1+p_2-2k)^\sigma\nonumber\\
&\qquad\qquad\qquad\qquad\qquad\qquad\qquad\qquad\times\frac{1}{(k-p_2)^2-M_{S_1}^2}\{-(\lambda_L^{ij})^*P_R+(\lambda_R^{ij})^*P_L\}\Bigg],\nonumber\\
&\equiv iQ_{\rm EM}^{S_1}\,N_C\int\frac{d^4k}{(2\pi)^4}\Bigg[\frac{\mathcal{N}_2^\sigma}{\mathcal{D}_2}\Bigg].
\label{eq:t2}
\end{align}
Here $Q_{\rm EM}^{S_1}=1/3$ is the EM charge of $S_1$. Recasting the numerator of Eq.~\eqref{eq:t2}, one gets,
\begin{align}
\mathcal{N}_2^\sigma=\frac{1}{2}\Big[\{\mathcal{A}_1\slashed{k}+\mathcal{A}_2 m_i\}+\gamma^5\{\mathcal{A}_3\slashed{k}+\mathcal{A}_4 m_i\}\Big] (p_1+p_2-2k)^\sigma.
\label{eq:lepg2_neu2}
\end{align}
Feynman parametrization recasts the denominator as, 
\begin{align}
\mathcal{D}_2&=(k-yp_1-zp_2)^2-M_{S_1}^2\Big[x\rho_i+(1-x)\Big]\nonumber\\
    &=n^2-\Delta_2(x).
\end{align}
Thus, the NP contributions to the anomalous magnetic moment and EDM, arising from the Type-2 diagram, can be formulated as,
\begin{align}
\label{eq:a2}
\Delta a^\ell_2&=-\frac{1}{16\pi^2}\Bigg[\mathcal{A}_1\left(\frac{m_\ell}{M_{S_1}}\right)^2G_3(\rho_i)+\mathcal{A}_2\left(\frac{m_\ell\,m_i}{M^2_{S_1}}\right)G_4(\rho_i)\Bigg],\\
d_2^\ell&=-\frac{e}{16\pi^2}\Bigg[\mathcal{A}_3\left(\frac{m_\ell}{M^2_{S_1}}\right)G_3(\rho_i)+\mathcal{A}_4\left(\frac{m_i}{M^2_{S_1}}\right)G_4(\rho_i)\Bigg],
\label{eq:d2}
\end{align}
where,
\begin{align}
G_3(w)&=\int_0^1\left[\frac{x(1-x)^2}{xw+(1-x)}\right]\,dx=\frac{1-6w+3w^2+2w^3-6w^2\,{\rm ln}\,w}{6(1-w)^4},\nonumber\\
G_4(w)&=\int_0^1\left[\frac{x(1-x)}{xw+(1-x)}\right]\,dx=\frac{1 - w^2 + 2w\,{\rm ln}\,w}{2 (1 - w)^3}\,.
\end{align}
Therefore, within this minimally extended BSM framework, the complete NP contribution to the leptonic observables can be defined as,
\begin{align}
\Delta a_\ell=\Delta a^\ell_1+\Delta a^\ell_2\,,\qquad |d_\ell|=\left|d_1^\ell+d_2^\ell\right|.
\label{eq:main}
\end{align}
\section{Numerical Analysis and Results}
\label{sec:num}
\noindent
In this section, we shall try to identify the allowed region of the parameter space through flavor-specific constraints. $\Delta a_\ell$ and the experimental upper bound on EDM will be considered simultaneously as the constraining factors for each generation. For completeness, one can enlist the free parameters of this model as follows:
\begin{align*}
\Big\{M_{S_1},\, \lambda_{L,R}^{u,\tau},\,\lambda_{L,R}^{c,\mu},\, \lambda_{L,R}^{t,e}\Big\}. 
\end{align*}
Note that, respecting the LHC constraints at $\sqrt{s}=13$ TeV one has to choose $M_{S_1}\geq 1.5$ TeV~\cite{CMS:2018ncu, ATLAS:2020dsk, ATLAS:2019ebv, CMS:2018lab, CMS:2018svy}. However, the NP couplings can be varied freely within the bounds of perturbative unitarity~\cite{Allwicher:2021rtd}. 
\begin{figure}[!ht]
\centering
\subfloat[(a)]{\includegraphics[scale=0.6]{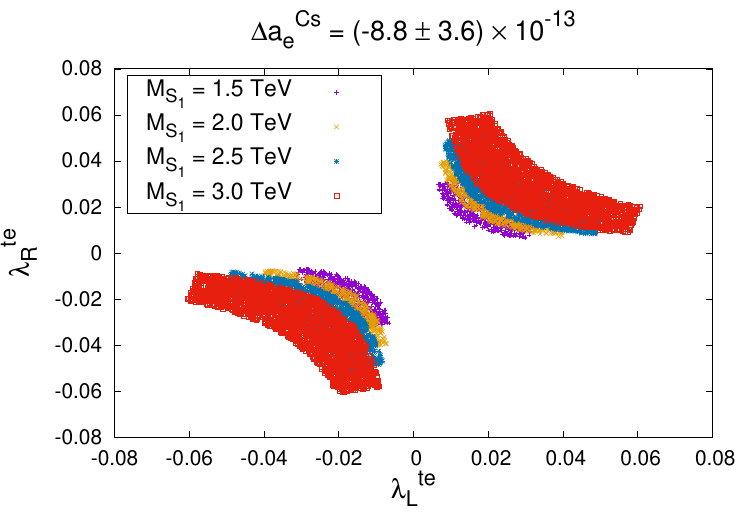}}\qquad\quad
\subfloat[(b)]{\includegraphics[scale=0.6]{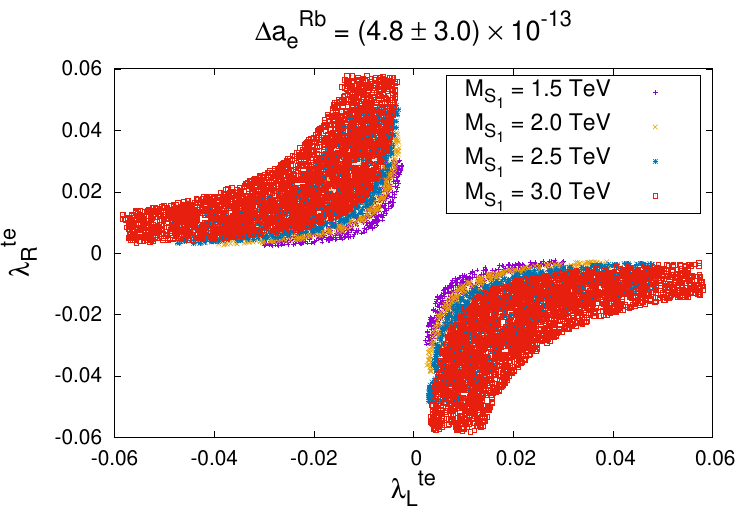}}\\
\subfloat[(c)]{\includegraphics[scale=0.6]{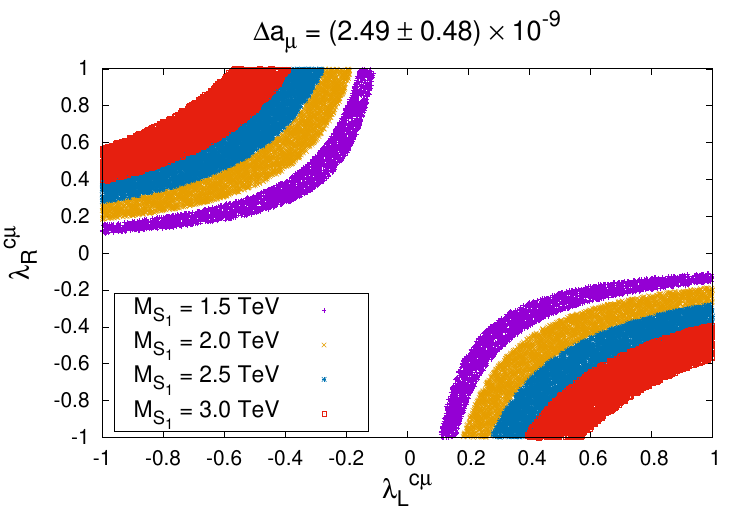}}
\caption{Allowed parameter space for the first two lepton generations constrained through (a) $\Delta a_e^{\rm (Cs)}$ \& $|d_e|$, (b) $\Delta a_e^{\rm (Rb)}$ \& $|d_e|$, and (c) $\Delta a_\mu$ \& $|d_\mu|$. The four different colors correspond to $M_{S_1}=1.5$ TeV~(violet), 2.0 TeV~(golden), 2.5 TeV~(sky blue), and 3.0 TeV~(red)}
\label{fig:para}
\end{figure}
Fig.~\ref{fig:para}(a) shows the allowed parameter space in the $\lambda_{L}^{t,e}-\lambda_{R}^{t,e}$ plain for a set of four $M_{S_1}$ values: $M_{S_1}=1.5$ TeV~(violet), 2.0 TeV~(golden), 2.5 TeV~(sky blue), and 3.0 TeV~(red). The depicted region simultaneously satisfies the observed $\Delta a^{\rm (Cs)}_e$ value and the experimental bound on $|d_e|$. However, it is a notable feature of this considered framework that even if one assumes the $\Delta a^{\rm (Rb)}_e$ results instead of Cs, a valid parameter space can be obtained~[see Fig.~\ref{fig:para}(b)]. Similarly, for the muon sector $\lambda_{L}^{c,\mu}-\lambda_{R}^{c,\mu}$ plain has been constrained through the $\mu$-specific observables, i.e., $\Delta a_\mu$ and $|d_\mu|$~[see Fig.~\ref{fig:para}(c)]. Note that, numerically, the same exercise can be repeated for the $\tau$-sector to constrain the $\lambda_{L}^{u,\tau}-\lambda_{R}^{u,\tau}$ region. However, the present experimental sensitivity is inadequate to probe the NP effects to $a_\tau$ and $d_\tau$ that one can obtain from Fig.~\ref{fig:loops}. Thus, no significant conclusion can be drawn in this case and the entire parameter space is effectively available.

Fig.~\ref{fig:para} leads to two interesting observations:
\begin{itemize}
\item {\it With increasing $M_{S_1}$ value, the magnitude of the couplings shifts to the higher side}. This particular behavior can be understood by analyzing the $M_{S_1}$-dependence of $\Delta a_\ell$ and $d_\ell$ for a fixed set of fermion masses. From Eqs.~\eqref{eq:a1}-\eqref{eq:d1} and \eqref{eq:a2}-\eqref{eq:d2} it is clear that there is an overall $M_{S_1}^2$ suppression. However, the complete $M_{S_1}$-dependence can only be noted by studying the individual variation of the functions $G_{\{1,2,3,4\}}$. Fig.~\ref{fig:fun} shows the variation of the $G$ functions with respect to $M_{S_1}$. For illustration, $m_\ell=m_\mu=0.105$ GeV and $m_i=m_c=1.275$ GeV have been assumed~\cite{ParticleDataGroup:2022pth}.
\begin{figure}[!ht]
\centering
\includegraphics[scale=0.6]{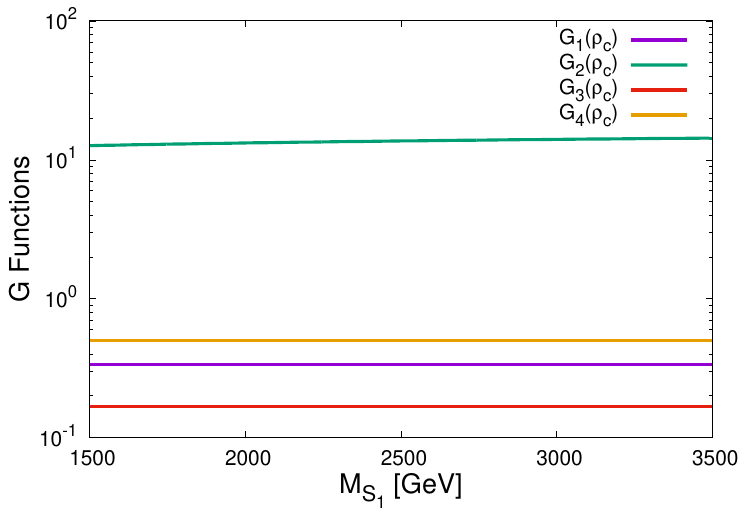}
\caption{Variation of $G_1,\,G_2,\,G_3$ and $G_4$ as a function of $M_{S_1}$ for  $m_\ell=m_\mu=0.105$ GeV and $m_i=m_c=1.275$ GeV. Here, $\rho_c=(m_c/M_{S_1})^2$.}
\label{fig:fun}
\end{figure}
Fig.~\ref{fig:fun} clearly indicates that $G_1,\,G_3,\,G_4$ do not exhibit any notable variation with the increasing $M_{S_1}$ value, while $G_2$ shows only a slight increment. Thus, to a good approximation, one can conclude that for a given set of quark and lepton masses, $\Delta a_\ell$ and $|d_\ell|$ decreases quadratically with $M_{S_1}$. Therefore, for compensating this suppression, the couplings must rise to match the experimental observations. 
\item {\it In Fig.~\ref{fig:para}(a) the product $\lambda_{L}^{t,e}\times\lambda_{R}^{t,e}$ is positive, whereas it flips to a negative value in Fig.~\ref{fig:para}(b).} This is a direct consequence of the oppositely aligned values of $\Delta a^{\rm (Cs)}_e$ and $\Delta a^{\rm (Rb)}_e$. From Fig.~\ref{fig:fun}, one can see that the function $G_2$ produces the leading contribution over the entire parameter space. The effect is further enhanced due to the chosen flavor ansatz~[see Eq.~\eqref{eq:flav}] as it connects the lightest lepton with the heaviest quark and vice versa. Thus, the sign of the term $\mathcal{A}_2\left(\frac{m_\ell\,m_i}{M^2_{S_1}}\right)G_2(\rho_i)$~[see Eq.~\eqref{eq:a1}], or to be more specific, the sign of $\mathcal{A}_2$ effectively decides the sign of $\Delta a_e$ in the theory. The same argument is valid for the negative values of $\lambda_{L}^{c,\mu}\times\lambda_{R}^{c,\mu}$ in Fig.~\ref{fig:para}(c).   
\end{itemize}  
\section{Conclusions}
\label{sec:conc}
\begin{justify}
This paper has considered a minimal extension of the Standard Model with a TeV-scale scalar Leptoquark $S_1$ transforming as $(\mathbf{\bar{3}},\,\mathbf{1},\,1/3)$ under the SM gauge group. In the presence of $S_1$, there can be corrections to the $\ell\ell\gamma$ vertex at the one-loop level, which may lead to new physics contributions to the lepton $(g-2)$ and EDM. A particular flavor structure has been chosen to suppress the CLFV processes while enhancing the BSM contributions to other low-energy lepton phenomena. The new one-loop contributions have been computed analytically, followed by a numerical scan to determine the parameter space allowed under the recent $(g-2)_\ell$ and EDM constraints for each of the lepton generations. Four different LQ masses have been considered to understand the phenomenological implication of the NP scale on flavor-specific low-energy observables. For the electron sector, viable parameter spaces have been found corresponding to both of the experimental results, i.e., $\Delta a^{\rm (Cs)}_e$~[see Eq.~\eqref{eq:e_Cs}] and $\Delta a^{\rm (Rb)}_e$~[see Eq.~\eqref{eq:e_Rb}]. Note that it is a significant feature of this work that it can explain both positive~($\Delta a^{\rm (Rb)}_e$ \& $\Delta a_\mu$) and negative~($\Delta a^{\rm (Cs)}_e$) discrepancies in the anomalous magnetic moment of leptons by simply rotating the parameter space while keeping the entire scenario consistent with the respective EDM discovery limits. Though the $\tau$-sector has also been analyzed but due to lower experimental sensistivity the complete parameter space is allowed within the perturbative bounds. However, the assumed model structure can explain any future update on $a_\tau$ and/or $d_\tau$ which can probe the BSM contributions to the $\tau$ phenomenology. Collider-based experiments searching for the TeV-scale scalar LQs and/or any experimental update on the low-energy lepton phenomena can be used to test or falsify the proposed framework. 
\end{justify}
%%%%%%%%%%%%%%%%%%%%%%%%%%%%%%%%%%%%%%%%%%%%%%%%%%%%%%%%%%%%%%%%%%%%%%%%%%%%%%%%%%%%%%    
\bigskip
\small \bibliography{Lepton_LQ_V2}{}
\bibliographystyle{JHEPCust}

\end{document}